# The 0.7-anomaly in quantum point contact; many-body or single-electron effect?

**Tadeusz Figielski**

Institute of Physics, Polish Academy of Sciences, Warsaw, Poland
E-mail: figiel@ifpan.edu.pl

Apart from usual quantization steps on the ballistic conductance of quasi-one-dimensional conductor, an additional plateau-like feature appears at a fraction of about 0.7 below the first conductance step in GaAs-based quantum point contacts (QPCs). Despite a tremendous amount of research on this anomalous feature, its origin remains still unclear. Here, an unique model of this anomaly is proposed relied on fundamental principles of quantum mechanics. It is noticed that just after opening a quasi-1D conducting channel in the QPC a single electron travels the channel at a time, and such electron can be – in principle – observed. The act of observation destroys superposition of spin states, in which the electron otherwise exists, and this suppresses their quantum interference. It is shown that the QPC-conductance is then reduced by a factor of 0.74. "Visibility" of electron is enhanced if the electron spends some time in the channel due to resonant transmission. Electron's resonance can also explain an unusual temperature behavior of the anomaly as well as its recently discovered feature: oscillatory modulation as a function of the channel length and electrostatic potential. A recipe for experimental verification of the model is given.

PACS numbers: 73.63.-b; 73.63.

## 1. Introduction

Short, narrow constrictions connecting two reservoirs of two-dimensional electron gas (2DEG) in semiconductor heterostructures, called quantum point contacts, exhibit at low temperature a quantization of the ballistic conductance in units of $G_0 = 2e^2/h$ ($e$ and $h$ are the elementary charge and the Planck's constant, respectively, and the factor 2 arises from spin degeneracy). The constriction behaves like a quasi-one-dimensional conducting channel. The channel is defined by a voltage applied to a pair of finger-like gates deposited on the top of heterostructure. Upon widening the channel by tuning the gate voltage, one observes a staircase increase in the conductance, displaying distinct plateaus at integer multiples of $G_0$ [1]. The plateaus arise from almost perfect electron transmission through 1D-energy subbands of the channel, each contributing a quantity $G_0$ to the conductance, whose number increases with the channel width.

Surprisingly, in clean channels of GaAs an additional plateau-like feature at about $0.7 G_0$ appears, which is commonly known as the 0.7 anomaly [2-4]. Upon applying a magnetic field parallel to the channel, the 0.7 feature evolves into the spin-split plateau at $e^2/h$, which reveals its relation to the electron spin. In contrast to other plateaus, the 0.7 feature becomes less pronounced at lower temperatures, evolving from a distinct "plateau" at 4.2 K to a vague shoulder at 20 mK. At sufficiently low temperature the 0.7 feature is accompanied by the so-called zero bias anomaly (ZBA): appearing a maximum in nonlinear conductance while the bias voltage is swept through zero. That ZBA is characteristic of the Kondo effect in quantum dots.

Recently, structures supplied with three pairs of finger-like gates were studied, which allowed tuning the QPC length. It has been found in those structures that the 0.7 feature exhibits oscillatory modulation as a function of the channel length [5]. Similar modulation has been observed while scanning negatively charged tip above the constriction surface [6]. It has been recently demonstrated that the 0.7 anomaly is not disturbed by the presence of defects localized in close proximity of the constriction [7]. Statistical study performed on 36 constriction units fabricated on the same wafer, and processed in the same way, have found a quantity $0.75G_0$ as the mean value of the conductance anomaly [8]. The 0.7 anomaly is also observed in *p*-type QPCs [9].

It is commonly believed that the anomaly is a many-body effect, and accordingly numerous explanations have been proposed [2,5,6,10,11]. Currently, Kondo-like effects are mostly invoked for those explanation [5,12,13]. However, more and more studies imply the Kondo effect is not linked to 0.7 feature [14-17]. So, two decades after the discovery, origin of this anomaly is still extensively debated. In this paper a unique single-electron model of the anomaly is proposed, relied on quantum-mechanical superposition and interference of spin states. The merit of this model is its simplicity and generality. It is the only model which predicts a concrete value for the fraction of $G_0$ characteristic of the anomaly and can explain all its main features.

## 2. Superposition and interference of the electron-spin states

Any spin state of the electron, $|\chi_i\rangle$, can be represented as a linear combination of two basic states, "spin up" $|\uparrow\rangle$ and "spin down" $|\downarrow\rangle$, with coefficients $a_i$ and $b_i$ that are complex numbers: $|\chi_i\rangle = a_i |\uparrow\rangle + b_i |\downarrow\rangle$, where $|a_i|^2 + |b_i|^2 = 1$. In the Riemann (Bloch) sphere representation the state $|\chi_i\rangle$ is depicted as a point on the surface of this sphere. In the spherical coordinate system it can be represented as a spinor $|\chi_i\rangle = \begin{pmatrix} a_i \\ b_i \end{pmatrix}$, where $a_i = \cos(\Theta_i/2) \exp(-i\Phi_i/2)$, $b_i = \sin(\Theta_i/2) \exp(i\Phi_i/2)$, and $\Theta$ and $\Phi$ denote the polar and azimuthal angle, respectively.

Principle of quantum superposition claims that any physical system – such as electron – exists partly in all its possible states simultaneously, as long as it is not being observed. The state of linear superposition can here be written as $|X\rangle = \sum_i c_i |\chi_i\rangle$, where the coefficients $c_i$ define contributions of different spin states to the superposition. In the absence of a magnetic field all spin states of the electron are equally probable and then $|X\rangle = C \sum_i |\chi_i\rangle$, where the summation runs over all possible spin states, and $C$ is normalization factor. Probability that the electron finds itself in that state is

$$P_S = |C|^2 \langle X|X\rangle = |C|^2 \left( \sum_i \langle \chi_i|\chi_i\rangle + \sum_{ij; j \neq i} \langle \chi_i|\chi_j\rangle \right). \tag{1}$$

The second sum in parenthesis results from quantum interference between different spin states. Because of infinite number of those states, the discrete values $\Theta_i$ and $\Phi_i$ should be replaced by continuous variables, and the summations – by integration over the surface of the unit sphere, $S$, according to a transformation

$$\sum_i f(\Theta_i, \Phi_i) \to \frac{1}{S} \oiint f(\Theta, \Phi) \mathrm{d}s = \frac{1}{4\pi} \int_0^\pi \int_0^{2\pi} f(\Theta, \Phi) \sin(\Theta) \mathrm{d}\Phi \mathrm{d}\Theta. \tag{2}$$

Using the spinor representation of different spin states, we find after the integration

$$P_S = |C|^2 (1 + I), \tag{3}$$

where the calculated interference term is $I \cong 0.36$.

Similar procedure can also be applied in case of spin polarization. Then, however, the coefficients $c_i$ appearing in the superposition are different, resulting in a definite degree of spin polarization, given by a ratio $P = \sum_i |c_i|^2 (|a_i|^2 - |b_i|^2) / \sum_i |c_i|^2$.

## 3. Suppression of interference by observation

If an individual electron is being observed (detected), information is extracted that it is no more in the state of superposition. After the observation, electron must find itself in a definite spin state although we do not know in which one. The act of observation destroys interference, like that in the canonical double-slit experiment. Then, to get the probability, one has to sum partial probabilities of individual states instead of their probability amplitudes. The ratio, $\kappa$, of the probability of finding electron when it is subject to observation, $P_O$, to that when it exists in the state of superposition, $P_S$, is $\kappa = P_O/P_S \cong 0.74$.

Consider an electron travelling a QPC via the lowest 1D energy subband. The electron can exist there either in the state of superposition or – if it is being observed – in one of the possible spin states; the latter excludes interference. Most importantly, it is not necessary to perform any real observation of the electron to suppress the interference. As demonstrated in the double-slit experiments, it is enough to create experimental conditions allowing such observation (see e.g. [18]), which is one of mysteries of quantum mechanics

Essential condition allowing observation of an individual electron in the QPC is that no more than one electron travels the constriction region at the same time [19], which occurs just after opening an 1D conducting channel. In order to detect the electron one could exploit an electrostatic coupling between the channel and the gate electrodes defining constriction. Electron entering the channel induces a positive charge on the gates which generates an additional voltage on the gate-channel capacitor. Virtual detection of a voltage pulse in an external circuit (supplying the gate voltage) proves that electron has entered the channel. The charging time, given here by $\tau_{ch} = \varepsilon/\sigma$, where $\varepsilon$ is the electrical permittivity and $\sigma$ is the conductivity of gate electrode, is rders of magnitude shorter than the electron's transit time through the channel. Moreover, detectability of an electron passing through the channel can be enhanced if the electron is trapped in the channel for some time.

## 4. Resonant transmission

In fact, the 1D channel in QPC can behave as a resonant cavity for the electron wave. In the ballistic regime the two-probe resistance of QPC stems entirely from "contact resistance" between 1D conducting channel and 2DEG reservoirs [20]. Sudden drops in potential at both ends of the channel induces partial reflection of the electron wave. Due to possibility of multiple reflections at both ends, we expect the transmission resonance, owing to which the electron is temporarily trapped inside the channel that behaves like a resonant delay line.

The width of quasi-1D channel in QPC varies with the position, $x$, along its length. Energy of the bottom of the lowest 1D subband, $\varepsilon_1(x)$, and the wave number of ballistic electron, $k(x)$, changes appropriately to the width's variation. Thus $\varepsilon_1(x)$ forms a smooth hill with a maximum at the constriction bottleneck. In the quasi-classical approximation, general solution of the Schrödinger equation for electron travelling within the channel will be

$$\psi(x) = \frac{C_1}{\sqrt{k(x)}} e^{i \int k(x) dx} + \frac{C_2}{\sqrt{k(x)}} e^{-i \int k(x) dx}, \tag{4}$$

where $k(x) = \sqrt{2m[E - \varepsilon_1(x)]}/\hbar$, $m$ is the electron effective mass, $E$ is energy, and $\hbar = h/2\pi$.

Condition of resonance is met when $\int_0^L k(x) dx = n\pi$, where $L$ is the channel length, and n is integer. For the sake of simplicity, we neglect further variation of the wave number along the channel by putting $\varepsilon_1 = \text{const}(x)$. It comes down the condition of resonance to the relation: $n\lambda = 2L$, where $\lambda$ is the wavelength of electron propagating through the channel. The wavelength of a ballistic electron coming out of 2DEG reservoir with the electrochemical potential $\mu$ is

$$\lambda = \frac{h}{\sqrt{2m(\mu - \varepsilon_1)}}. \tag{5}$$

The quantity $E_F = \mu - \varepsilon_1$ represents an effective Fermi energy in the channel. Assuming tentatively the channel length to be $L = 200$ nm, we find in resonance $E_F = 0.14$ meV, and the Fermi velocity $v_F = 2.7 \times 10^4$ m/s (this and further numerical calculations concern *n*-GaAs). In equilibrium this $E_F$ would determine the one-dimensional density of electrons which, under resonance condition, corresponded to two electrons in the channel.

## 5. Reduction in the conductance

Current considerations can be summarized as follows. Electron in a 2DEG reservoir, prior to its entry to QPC, finds itself in a state of superposition of all possible spin states. Probability of finding that electron, assumed the further to be unity, contains a contribution originating from interference between different spin states. Individual electron entering the 1D-channel can be effectively detected if it is trapped in the channel for some time owing to resonant transmission. Detection of the electron suppresses the interference between different spin states. In other words: detection of the electron means its interaction with environment (that constitute here the gates connected with an external circuit) which causes decoherence of the state of superposition. The interference term $I$ in Eq.(3) becomes then equal to zero. Hence, probabilities of finding the electron after and before its entry to the channel are different; their ratio is 0.74. Obviously, the probability of finding the electron anywhere in the structure must be conserved. Here, this requirement comes down to conserving continuity of the probability current at the boundaries between 2DEG reservoirs and 1D channel. It can be met only if the electron wave-packet entering the channel is partly scattered back to the reservoir. That back-scattering contributes to an additional "contact resistance" which reduces the conductance of QPC just by a factor of 0.74.

## 6. Modulation of the anomaly

Consider now the intriguing effect of modulation of 0.7 anomaly by the electrostatic potential, reported in [6]. We attribute this modulation to repeatable occurrence of the electron's resonance in 1D channel. Resonance in a cavity occurs when the wave returning to its starting point – after reflection from the back wall of the cavity – meets the wave just starting in the same phase. So, during travelling back and forth the wave has to acquire a phase $\varphi = 2\pi$, or multiple of this value. This requirement leads directly to the condition of resonance: n $\lambda = 2L$.

When a negative electrostatic potential, $U$, is imposed on the QPC, the subband-edge energy, $\varepsilon_1$, is lifted up shutting the channel. In order to again open the channel, one has to adjust the gate voltage (making it less negative) to compensate the *U*-induced shift by widening the channel which lowers $\varepsilon_1$. However, by imposing the potential $U$ an additional phase $\Delta\varphi = -eU\tau/\hbar$ is acquired by the electron wave (like that in the electrostatic Aharonov-Bohm effect), where $\tau$ is the time of travelling the channel back and forth. We have $\tau = 2L/v$, where $v = \hbar k/m$ stands for the group velocity. The condition of resonance is restored when $\Delta\varphi = 2emUL/\hbar^2 k = 2\pi$ (here, an effect of gate voltage on $U$ has been neglected). While tuning electrostatic potential in the channel the resonance condition appears repeatedly with a period

$$\Delta U = \frac{h^2}{4emL^2}. \tag{6}$$

This relationship predicts just the period of modulation of the anomaly. For $L = 200$ nm we find $\Delta U = 1.1$ mV. This is a reasonable value to account for the observed modulation of the 0.7 anomaly while scanning negatively charged tip above the QPC [6].

In [5], an oscillatory modulation of the anomaly by tuning the channel length was observed on three-pairs-gate devices. Within the present model, origin of this modulation is similar to that discussed previously and can be described by Eq.(6), taking into account that $U$ is now generated by gate voltages. In the cited work the channel length was tuned continuously by changing the ratio of the voltage applied to the outer pair of gate electrodes, $V_{g2}$, to that applied to the central ones, $V_{g1}$. All these gate voltages contribute to the electrostatic potential acting on electrons in the channel. Their contribution manifests itself as more and more less negative gate voltage, $V_{g1}$, required for opening the channel while the ratio $V_{g2}/V_{g1}$ is being increased. In conclusion, we propose that tuning the electrostatic potential in the channel, and not its length, is the primary reason for the modulation of 0.7 anomaly.

## 7. Temperature dependence

The 0.7 anomaly appears in the range of electron energies (that translates into a range of gate voltages) in which the resonance of single electron in the channel enhances its detectability. Accordingly, an extension of the 0.7 feature on the gate-voltage scale would be determined by the resonance linewidth. Here, we assume tentatively that the dominant mechanism of damping resonance is dephasing of the wave function. The phase-coherence length of electron, $L_\varphi$, is a material-related parameter that decreases with temperature (approximately as $L_\varphi \propto T^{-1/3}$), which causes that the extension of the 0.7 "plateau" becomes less pronounced at lower temperatures.

Consider this issue in more detail. Uncertainty principle between momentum, $p$, and position, $x$, claims that $\Delta p \Delta x \geq \hbar/2$. Taking into account that $E = p^2/2m$, and putting $\Delta x = L_\varphi$, we find a relationship

$$\Delta E \geq \frac{E_F}{\pi} \frac{L}{L_\varphi}, \qquad (7)$$

where $E_F$ is the effective Fermi energy in resonance. This relation determines the spread of electron energies, $\Delta E$, within the 0.7 anomaly appears. To have an idea about magnitudes of the quantities considered here, we put into Eq.(7) the values of $L_\varphi$ obtained for 2DEG in GaAs/GaAlAs heterojunction [21]. The phase-coherence length is $L_\varphi = 3$ μm at temperature of 25 mK and falls down to about 0.6 μm at 1.3 K. Using these values one finds from Eq.(7) for $L = 200$ nm: $\Delta E = 3$ μeV and $\Delta E = 15$ μeV, respectively.

## 8. *P*-type channels

The case of holes in the valence band of GaAs is much more complex than that of electrons in the conduction band, mainly because of a strong spin-orbit interaction. Holes passing through *p*-type QPC just after opening the conducting channel are the heavy holes with relatively small wave numbers. Those holes behave as particles with effective spin $J = 3/2$, which have projections on the quantization axis $J_z = \pm 3/2$. Due to a size quantization, the holes coming out from a 2D reservoir have a quantization axis oriented in the epitaxial-growth direction (*z*-axis). It has been shown by Majorana [22,23], that a spin-n/2 particle can be represented as a set of n spin-1/2 particles. This finding enables us to calculate the spin-interference term for holes being in state of superposition. At this aim we can simply use the same spinor representation as for spin-1/2 particle. However, the values of coefficients $c_i$ in superposition are now diversified to favor the mean spin vector aligned along the quantization axis. Let us assume tentatively $c_i = \cos(\theta_i/2)$, which causes 17% spin polarization, and apply similar computational procedure as that in section 2. After the calculations we find a value of the interference term (equivalent to *I* in Eq.(3)) equal to 0.32, and hence $\kappa \cong 0.76$. This value is very close to $\kappa \cong 0.74$ obtained for electrons as a fraction characteristic of the anomaly. The ratio $\kappa$ turns out to be not very sensitive to the degree of spin polarization within a considerable range of the latter.

## 9. Conclusions

The model proposed here is able to explain all main features of the 0.7 anomaly. In particular, it predicts a concrete value of the anomalous conductance, which has never been attempted by previous theories. It has been assumed here, in accord with Ref.[14-17], that the 0.7 anomaly and the zero-bias anomaly are correlated but separate and distinct effects. Actually, the ZBA disappears above the temperature at which the 0.7 anomaly becomes the most distinctive. Recent phase-sensitive measurements on the QPC [24] suggest a connection between the ZBA and the Kondo effect.

This model can be verified experimentally. While tuning negative potential of the metallic tip placed above the QPC we expect to observe a periodic modulation of the anomaly whose period displays square dependence on the inverse channel length, described by Eq.(6).

**Acknowledgements**

Thanks are due to Wlodek Zawadzki for the critical reading the manuscript and useful advices. This work was partly supported by the National Science Centre (Poland) under Grant No 2011/03/B/ST3/02457

**References**

[1] van Wees B.J., van Houten H., Beenakker C.W.J., *et al.* (1988) Quantized Conductance of Point Contact in a Two-dimensional Electron Gas. *Phys. Rev. Lett.* **60**, 848-850.

[2] Thomas K.J., Nicolls J.T., Simmons M.Z., *et al*. (1996) Possible Polarization in One-dimensional Electron Gas. *Phys. Rev. Lett*. 77, 135-138.

[3] Thomas K.J., Nicolls J.T., Appleyard N.J., *et al*. (1998) Interaction Effects in One-dimensional Constriction. *Phys. Rev*. *B* 58, 4846-4851.

[4] Micolich A.P., *J. Phys.: Condens. Matter* (2011) What Lurks Below the Last Plateau: Experimental Studies of the 0.7 x $2e^2/h$ Conductance Anomaly in One-dimensional System. **23**, 443201.

[5] Iqbal M.J., Levy R., Koop E.J., *et al*. (2013) Odd and Even Kondo Effects from Emergent Localization in Quantum Point Contact. *Nature* **501**, 79-83.

[6] Brun B., Martins F., Faniel S., *et al*. (2014) Wigner and Kondo Physics in Quantum Point Contacts Revaled by Scanning Gate Microscopy. *Nature Commun*. **5**, 045426..

[7] Iagallo A., Paradiso N., Roddano S., *et al*. (2015) Scanning Gate Imaging of Quantum Point Contacts and the Origin of the 0.7 Anomaly. *Nano Research* **8**, 948-956.

[8] Smith L.W., Al-Taie H., Swigakis, F., et al. (2014) Statistical Study of Conductance Properties of One-dimensional Quantum wires Focusing on the 0.7 Anomaly. *Phys. Rev. B*, **90**, 045426.

[9] Komijani Y., Csontos M., Ihn T., *et al*. (2013) Origin of Conductance Anomalies in a *P*-type GaAs Quantum Point Contact. *Phys.Rev. B*, **87** 245406.

[10] Bauer F., Heyder J., Schubert E., *et al.* (2013) Microscopic Origin of the "0.7-anomaly" in Quantum Point Contact. *Nature* **501**, 73-78.

[11] Schubert E., Heyder J., Bauer F., *et al.* (2014) Toward Combined Transport and Optical Studies of the 0.7-anomaly in Quantum Point Contact. *Phys. Stat. Sol. B*, **251**, 1931-1937.

[12] Cronenwett S.M., Lynch H.J., Goldhaber-Gordon D. et al. (2002) Low-Temperature Fate of the 0.7 Structure in a Point Contact: a Kondo-like Correlated State in an Open System. *Phys. Rev. Lett*. **88**, 226805.

[13] Meir Y., Hirose K.Y., and Wingreen N.S. (2002) Kondo Model for the "0.7 Anomaly" in transport Through a Quantum Point Contact. *Phys. Rev. Lett*. **89**, 196802.

[14] Chen T.M., Graham A.C., Pepper M., *et al.* (2009) Non-Kondo Zero-bias Anomaly in Quantum Wires. *Phys. Rev. B*, **79**, 153303.

[15] Sfigakis F., Ford C.J.B., Pepper M., *et al*. Phys. (2008) Kondo Effect from a Tunable Bound State within a Quantum Wire. *Phys. Rev. Lett*. **100**, 026807.

[16] Xiang S., Xiao S., Fuji K., et al. (2014) On the Zero-bias Anomaly and Kondo Physics in Quantum Point Contact near Pinch-off. *J. Phys.: Condens. Matter* **26**, 125304.

[17] Kawamura M., Ono K., Stano P., *et al*. (2015) Electronic Magnetization of a Quantum Point Contact Measured by Nuclear Magnetic Resonance. *Phys. Rev. Lett*. **115**, 036601.

[18] Walborn S., Terra Cunha O., Pádua S., and Monken C.H. (2002) Double Slit Quantum Eraser. *Phys. Rev. A*, **65**, 033818.

[19] Figielski T. (2013) The puzzle of the 0.7 Anomaly in Conduction of Quantum Point Contact Reexamined. *Acta Phys. Polon. A*, **124**, 841-842.

[20] de Picciotto R., Stormer H.L., Pfeffer L.N., *et al*. (2001) Four Terminal Resistance of a Ballistic Quantum Wire. *Nature* **411**, 51-54.

[21] Ferrier M., Angers L., Rowe A.C.H., *et al*. (2004) Direct Measurements the Phase-coherence Length in a GaAs/GaAlAs Square Network. *Phys. Rev. Lett*. **93**, 246804.

[22] Majorana E. (1932) *Nuov. Cim*. **9**, 43.

[23] Penrose R., (1994) Shadows of the Mind, Oxford University Press, Oxford, chap.5.10.

[24] Brun B., Martins F., Faniel S., *et al*. (2016) Electron Phase Shift at the Zero-bias Anomaly of Quantum Point Contact. *Phys. Rev. Lett*. **116**, 136801.